\documentclass[5p,twocolumn]{elsarticle}

\usepackage{graphicx}
\usepackage{dcolumn}
\usepackage{latexsym}
\usepackage{amssymb}
\usepackage{bm}

\begin{document}
\begin{frontmatter}
\title{Noise reduction algorithm for Glueball correlators}
\author[a]{Pushan Majumdar,}
\ead{tppm@iacs.res.in}
\author[b]{Nilmani Mathur} 
\ead{nilmani@theory.tifr.res.in}
\author[a]{Sourav Mondal} 
\ead{tpsm5@iacs.res.in}

\address[a]{Department of Theoretical Physics,
Indian Association for the Cultivation of Science, Kolkata}
\address[b]{Department of Theoretical Physics,
Tata Institute of Fundamental Research, Mumbai}

\begin{abstract}
We present an error reduction method for obtaining 
glueball correlators from monte carlo simulations of SU(3) lattice 
gauge theory. We explore the scalar and tensor channels at three different 
lattice spacings. Using this method we can follow glueball 
correlators to temporal separations even up to 1 fermi. 
We estimate the improvement over the naive method and compare our results 
with existing computations.
\end{abstract}


\end{frontmatter}

\section{Introduction}
Glueballs are colourless states in Quantum Chromodynamics (QCD) made entirely out 
of gluons. QCD predicts the existence of glueballs. However, no glueball has yet 
been discovered unambiguously even though there are several candidate glueball resonances, 
such as $f_0(1370)$, $f_0(1500)$, $f_0(1710)$, $f_J(2220)$ etc.~\cite{pdg}. One 
reason is that glueball states can mix with mesons in the same $J^{PC}$ 
channel and so it is very difficult to unambiguously extract glueball masses
 experimentally. It remains nevertheless a very exciting proposition and for a 
recent review on the status of glueball searches we refer the reader to ref.~\cite{ochs}.

Glueball masses can be computed in lattice quantum chromodynamics and a lot of 
effort has been directed towards this computation. However there is still no 
consensus regarding the mass spectrum. It is a difficult computation in lattice 
QCD with dynamical fermions due to the high masses of the glueballs ($>$ 1 GeV) and their mixing with
mesonic operators in the same symmetry channels. In recent times computation of glueball masses  
in lattice QCD with dynamical fermions have been attempted in references~\cite{ferm1,ferm2,ferm3}.

Glueball masses are often computed in pure Yang-Mills theory. Advantages are that there 
is no mixing with mesonic operators and the glueballs are stable as they cannot decay. 
Thus it is much easier to extract 
the glueball masses from monte-carlo simulations of pure Yang-Mills theory than lattice 
QCD with dynamical fermions. Nevertheless, even in this theory, glueball correlation functions are  
 dominated by statistical noise at large temporal separations and contribution from excited 
states at short separations. Global fits become difficult and  
one often computes the ``effective mass" which is the logarithm 
of the ratio of the values of the correlation function between successive time slices.
If the effective mass does not vary over a significant temporal range then one assumes 
that the effective mass is the same as the globally fitted mass. 

To remove the effect of excited states, conventional methods involve 
computing correlation matrices with matrix elements between a large 
set of interpolating operators constructed from smeared or fuzzed links~\cite{fuzz} in the 
relevant symmetry channel\footnote{Computation of glueball correlators using the wilson flow 
for smoothening the gauge fields has been reported in~\cite{flow}.}. 
The ground state is obtained by diagonalizing the correlation matrix 
in each channel~\cite{ukqcd,VW,LTW}. As it is difficult to follow the correlator signal to 
large physical distances, even using the above techniques, one often uses asymmetric 
lattices with a significantly smaller temporal spacing compared to the spatial lattice 
spacing with the expectation to observe a flat behaviour of the effective masses~\cite{MornPear,nilmani}
 over several time slices.

A different approach is to use noise reduction algorithms. Such algorithms have been 
used in the past for computing the glueball spectrum 
for U(1), SU(2) and SU(3) lattice gauge theories~\cite{meyer1,meyer2,mkk,meyer3,MeyTep,dmgiu}. 

In this article we follow the latter approach. We restrict ourselves to pure Yang-Mills 
theory with gauge group SU(3) and employ only the standard operators 
in each $J^{PC}$ channel (scalar and tensor) but try to follow the correlator to large temporal 
separations using a new noise reduction algorithm. Since the contamination due to excited 
states fall off exponentially, we expect that correlators at distances beyond 0.5 fermi to 
be dominated by the ground state. 

In section 2 we describe the algorithm. Section 3 is devoted to 
our results on the correlators and masses for the scalar and tensor channels.
In section 4 we discuss the improvement obtained over existing conventional methods. Finally
in section 5 we draw our conclusions and outline directions for future studies.

\section{Algorithm}
\begin{table*}
\begin{center}
\begin{tabular}{|c|c|l|l|c|r|c|r|}
 \hline
Lattice & Size & ~~$\beta$ & ~~$({r_0}/{a})$& $^{\rm sub-lattice}_{\rm thickness}$ & iupd & loop size & \# meas.  \\ \hline
A & ${10^3}\times18$ & ~5.7  & ~2.922(9)  & 3 &  30 & $2\times 2$ & 1000000 \\ \hline
B & ${12^3}\times18$ & ~5.8  & ~3.673(5)  & 3 &  25 & $3\times 3$ & 1248000 \\ \hline
C & ${16^3}\times20$ & ~5.95 & ~4.898(12) & 4 &  50 & $5\times 5$ & 1024000 \\ \hline
\hline
D & ${12^3}\times18$ & ~5.8  & ~3.673(5)  & 3 &  70 & $3\times 3$ & 5760000 \\ \hline
E & ${12^3}\times20$ & ~5.95 & ~4.898(12) & 5 & 100 & $5\times 5$ & 3456000 \\ \hline
F & ${12^3}\times20$ & ~6.07 & ~6.033(17) & 5 & 100 & $5\times 5$ & 1536000 \\ \hline
\end{tabular}
\caption{\label{lattices} Simulation parameters for all the lattices. Lattices A,B and C were used
for the scalar channel while D,E and F were for the tensor channel.}
\end{center}
\end{table*}

We compute glueball correlators using Mote-Carlo simulations of SU(3) lattice gauge theory with the 
Wilson gauge action at three different lattice spacings for both the scalar and the tensor channels.
For updating the links we use the usual Cabibbo-Marinari heat-bath for SU(3) and use three over-relaxation
steps for every heat-bath step. We set the scale on the lattice through the Sommer parameter $r_0$~\cite{r0}.
Our simulation parameters are given in table~\ref{lattices}. The Sommer parameter for our lattices 
have been computed in~\cite{guasomwitt} and we use those values.

The noise reduction scheme we implement follows the philosophy of the multilevel algorithm.
The multilevel algorithm was introduced in \cite{ml} as an exponential noise reduction technique 
for measuring polyakov loop correlators in lattice gauge theories with a local action. 
However the principle is general and can be applied to other observables as well. In addition to 
polyakov loop correlators, it has been used to measure observables such as the polyakov loop~\cite{ploop}, 
wilson loop~\cite{wloop}, components of the energy-momentum tensor~\cite{emt} as well as the 
glueball mass spectrum~\cite{meyer1,meyer2,mkk,meyer3,MeyTep,dmgiu}.

\begin{figure}
\begin{center}
\includegraphics[width=8cm]{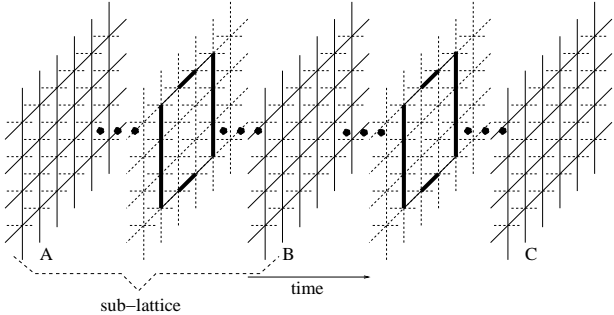}
\caption{\label{fig:scheme} Multilevel scheme for computing glueball correlators. The time slices marked
A, B and C are held fixed during the sub-lattice updates. The thick links are the ones which
are replaced by their multihit averages.}
\end{center}
\end{figure}

The main principle of the multilevel algorithm is to compute expectation values in a nested manner.
Intermediate values are first constructed by averaging over sub-lattices with boundaries and then
the full expectation values are obtained by averging over the intermediate values with different 
boundaries. We refer the reader to~\cite{mkk} for details.  

For our implementation, we slice the lattice along the temporal direction by fixing the spatial links and 
compute the intermediate expectation values of the glueball operators by performing several sub-lattice 
updates. Individual correlators are created using products of the averaged operators at different time 
slices. The scheme is depicted in figure \ref{fig:scheme} and the extents 
of the sub-lattices and number of sub-lattice updates along with other simulation parameters are 
shown in table~\ref{lattices}.  

The glueball operators between which we compute our correlation function (source and sink) are 
extended wilson loops denoted by $P_{ab}$ where $a,b$ go over the three spatial directions $x,y,z$. 
The operators are projected to zero momentum states as usual.
We denote the temporal separation between the source and sink operator by $\Delta t$.
The sizes of the loops used for the different lattices are given in table~\ref{lattices}.
Correlation functions between large loops have the advantage that they have much less 
contamination from the higher excited states compared to those between elementary plaquettes. 
Such an approach was reported in \cite{rgupta}. There, however, single exponential fits to the 
correlators were not possible as the data was too noisy. Nevertheless  
it was observed that glueballs seemed to have the largest overlap with loops 
of spatial extent 0.5 fermi in each direction. We therefore choose loops of roughly the extent 
$r_0\times r_0$ to construct our glueball operators.

Our first noise reduction step is a semi-analytic multihit on the SU(3) links \cite{su3mhit} 
with which the wilson loops are constructed 
and in addition we use sub-lattice updates to obtain the expectation values of the glueball
operators with very little noise. The choice of the number of sub-lattice updates ``iupd" is an 
important parameter of the algorithm. For the tensor channel, the rule of the thumb we follow is that 
the operator expectation 
value over the sub-lattice updates should be the same order as the square root of the correlator at 
a large value of $\Delta t$. For the scalar channel the same holds but for the connected parts. 
We compare the overall noise reduction of our algorithm with the naive method 
(where operators are constructed from elementary plaquettes and
only full lattice updates using heat-bath and over-relaxation is used) in section~\ref{adv}.

The multilevel algorithm is very efficient for calculating quantities with very small expectation 
values. While the operators in the tensor channel viz. ${\mathcal E}_1={\mathbb Re}\left ( P_{xz}
-P_{yz}\right )$ and ${\mathcal E}_2={\mathbb Re}\left ( P_{xz} + P_{yz} - 2 P_{xy} \right )$
have zero expectation values and are therefore ideal for direct evaluation using the multilevel 
scheme, the scalar operator ${\mathcal A}={\mathbb Re}\left ( P_{xy} + P_{xz} + P_{yz} \right )$
has a non-zero expectation value which has to be subtracted to obtain the connected correlator.
For the scalar channel, we therefore do the simulation in two steps. The first step is to determine the 
expectation value of the glueball operator. This has to be determined very accurately so that 
the error in the expectation value has negligible contribution to the error on the correlator. 
Otherwise the error on the expectation value of the operator will dominate the total 
error and further error reduction on the correlator would be impossible. We use multi-hit 
on the links to determine the expectation value of the glueball operator. While this was sufficient 
for our loop size and coupling, if necessary a multi-level 
scheme can also be used for this estimate. Then we directly computed the connected correlator using 
 $\left ( {\mathcal A} - \langle {\mathcal A}\rangle\right )$ as the 
operator with a zero expectation value. The choice of ``iupd" was done in the same way as 
in the tensor channel.  

An alternative to the above is to evaluate the derivative of the glueball correlator 
directly as that does not need a subtraction. This was carried out in \cite{mkk} for the U(1) case. 
In our current calculations we found that the subtraction procedure was more efficient compared to 
evaluating the derivative.

We observed one more phenomenon which is particular to this algorithm. For the smaller values of 
$\Delta t$ where most contributions come from slices which are within the same sub-lattice, there 
are strong effects due to the short temporal extent of the sub-lattice itself. In such cases we 
were forced to take into account only correlators between those time slices which lay in different 
sub-lattices. We found this effect to be significant only in the tensor channel (probably because of the 
larger value of ``iupd" in those cases).   

\section{Results - masses}
\begin{figure*}
\begin{center}
\includegraphics[width=5.5cm,angle=-90]{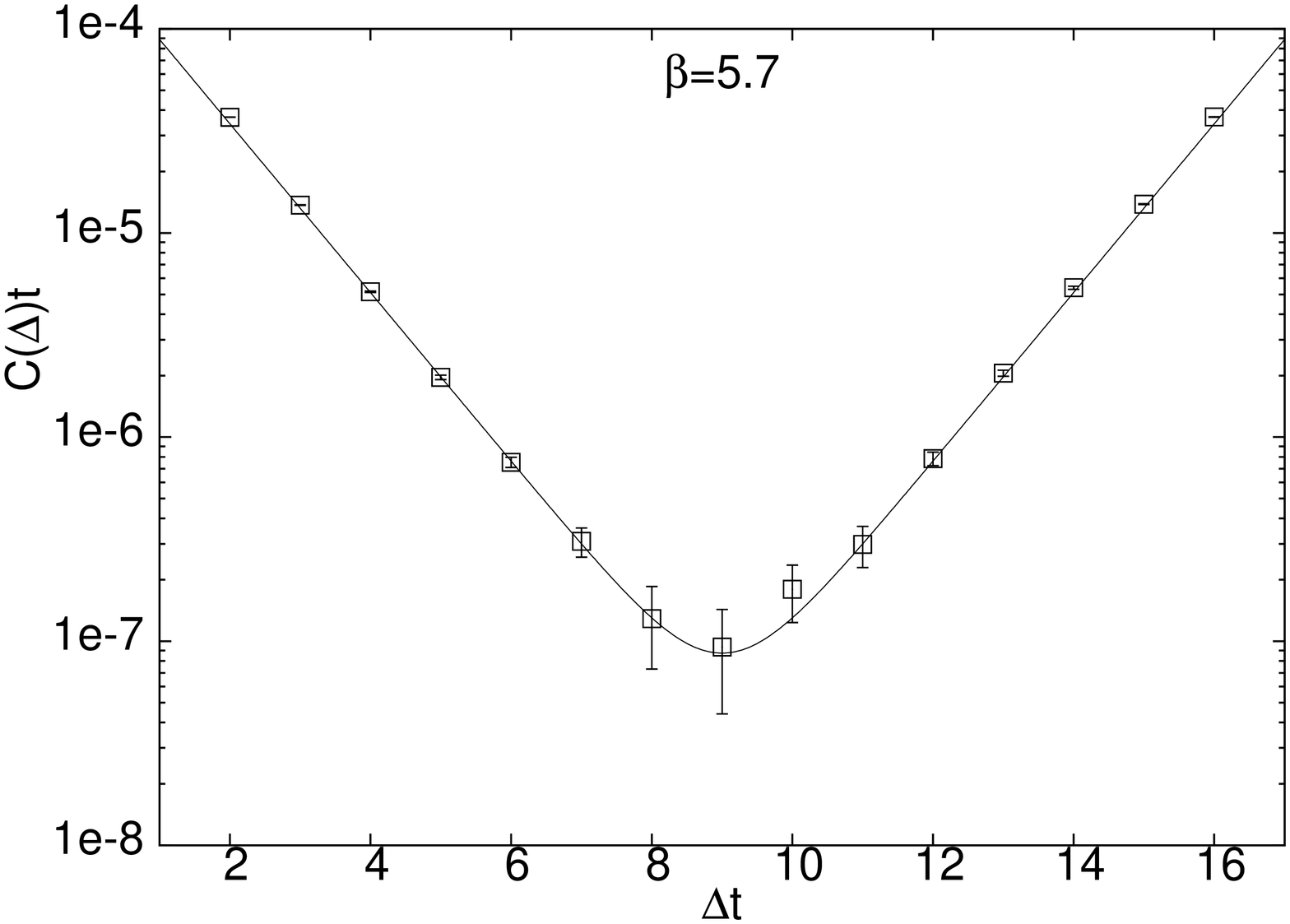}
\includegraphics[width=5.5cm,angle=-90]{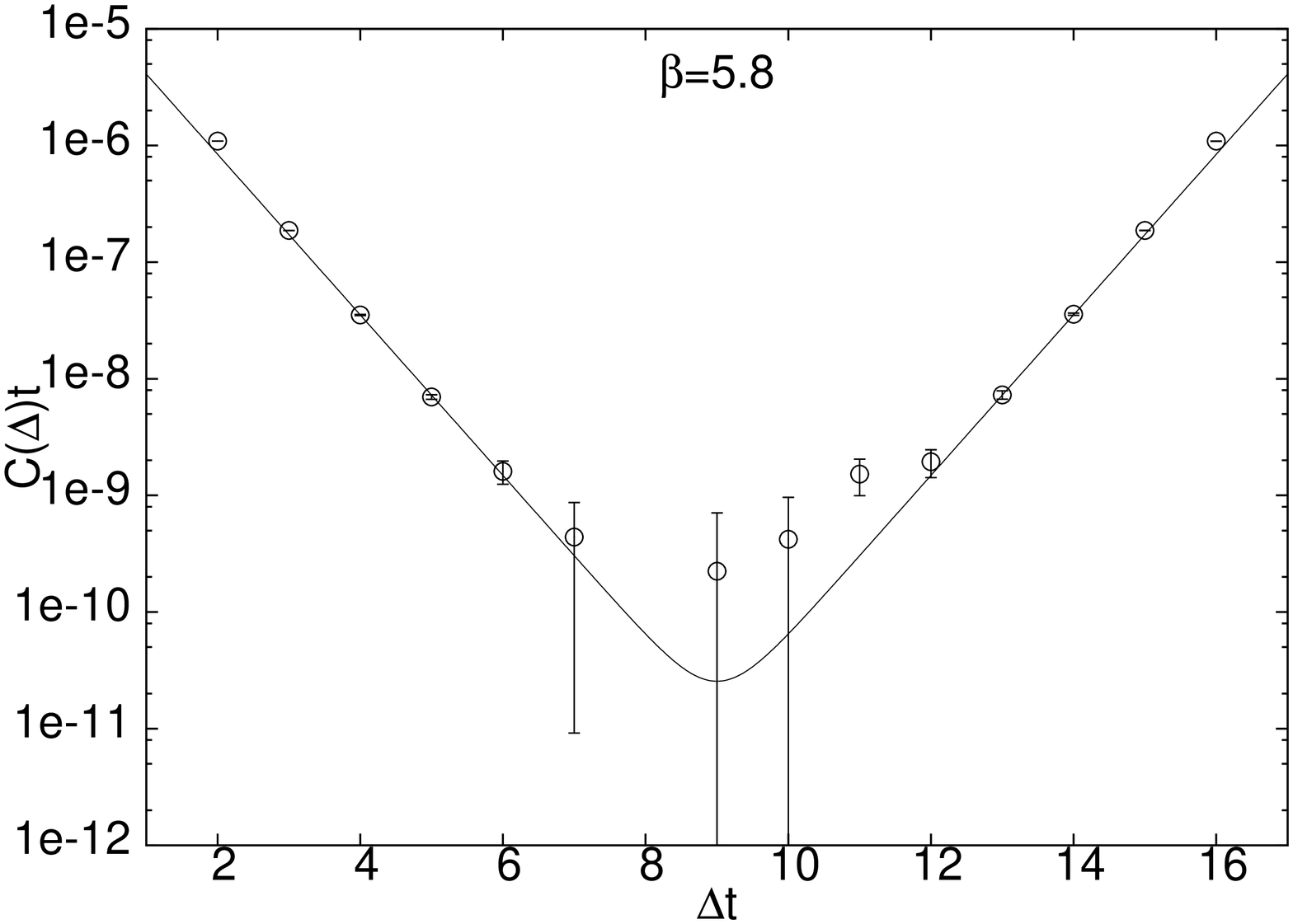}
\\
\includegraphics[width=5.5cm,angle=-90]{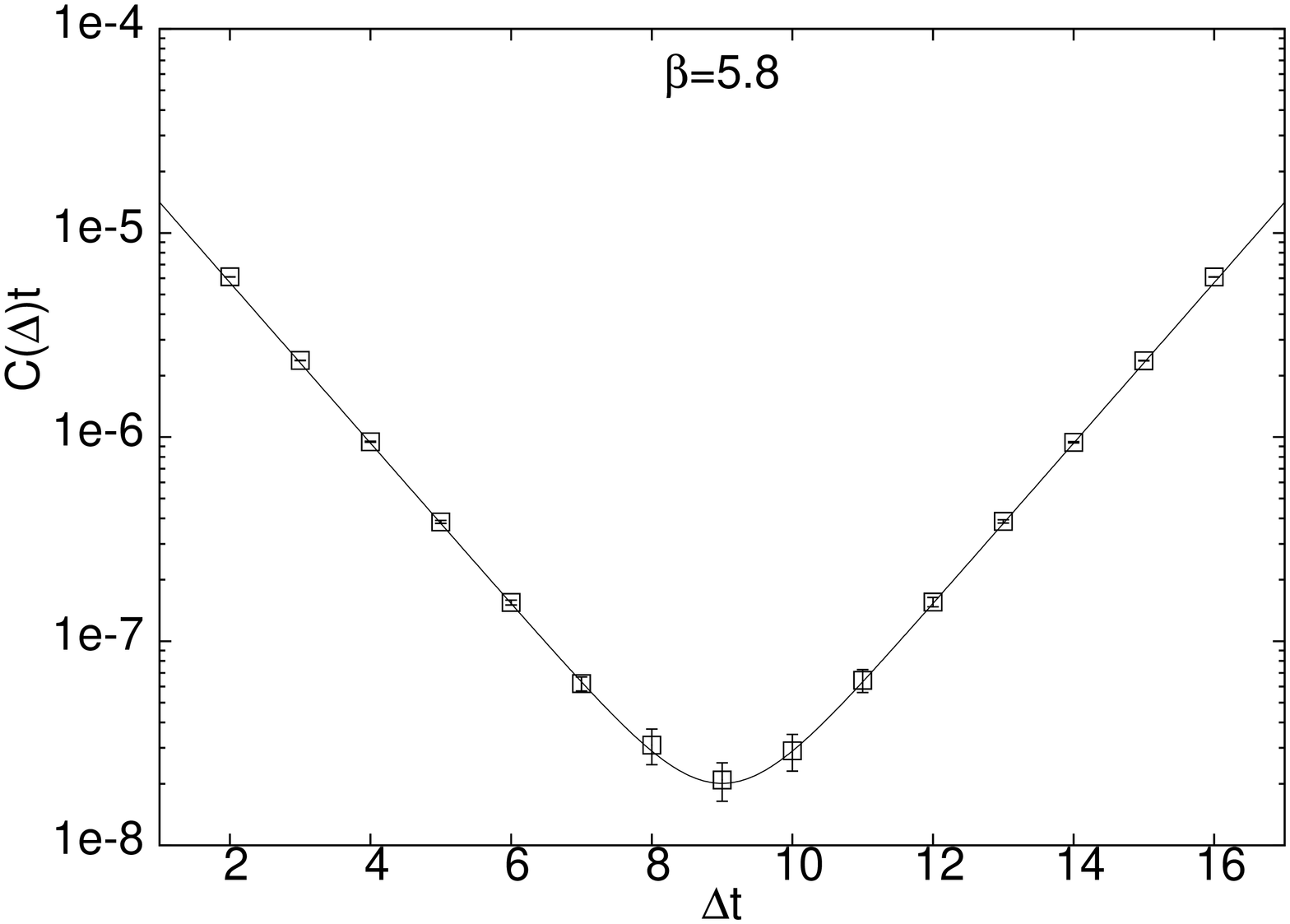}
\includegraphics[width=5.5cm,angle=-90]{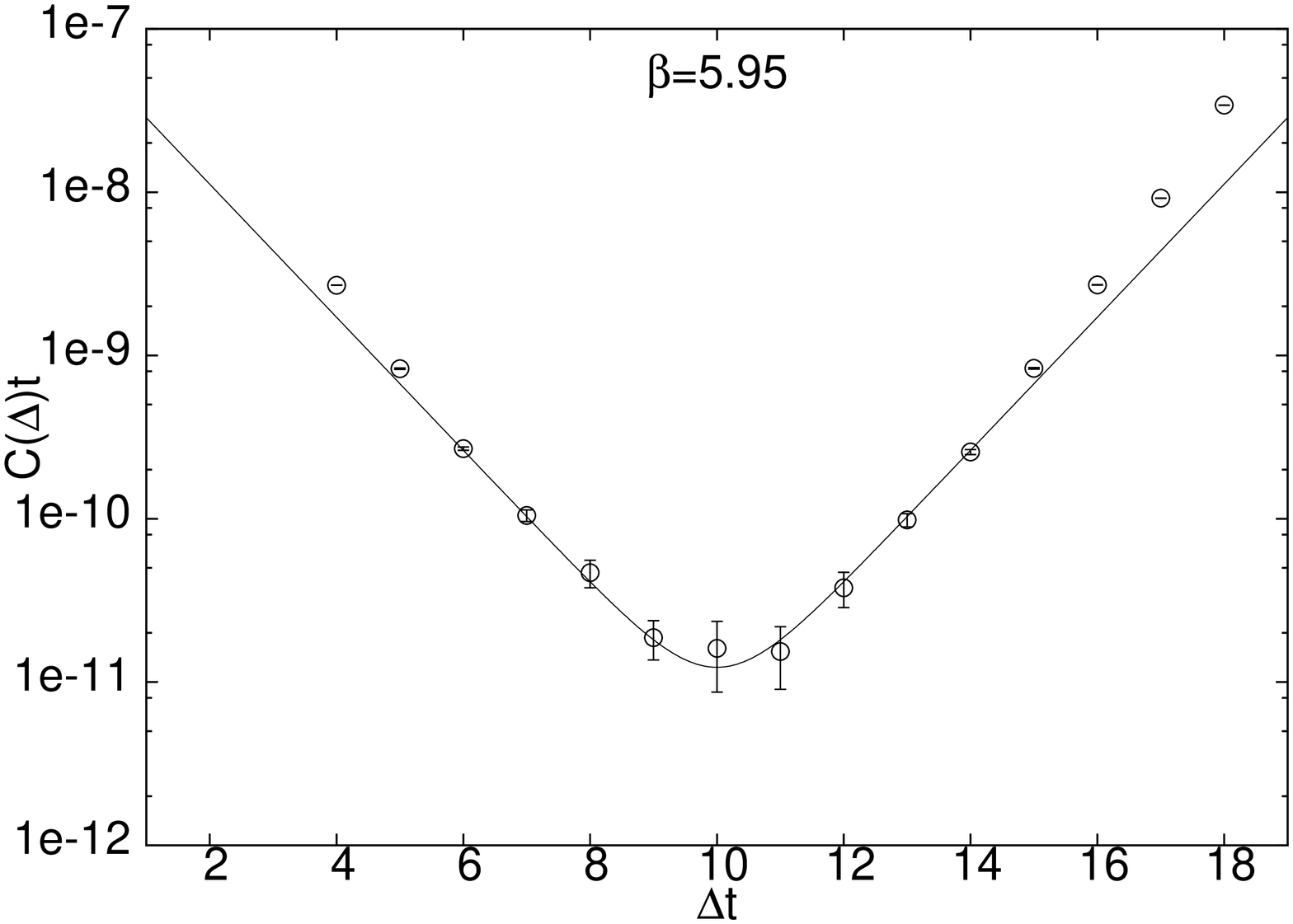}
\\
\includegraphics[width=5.5cm,angle=-90]{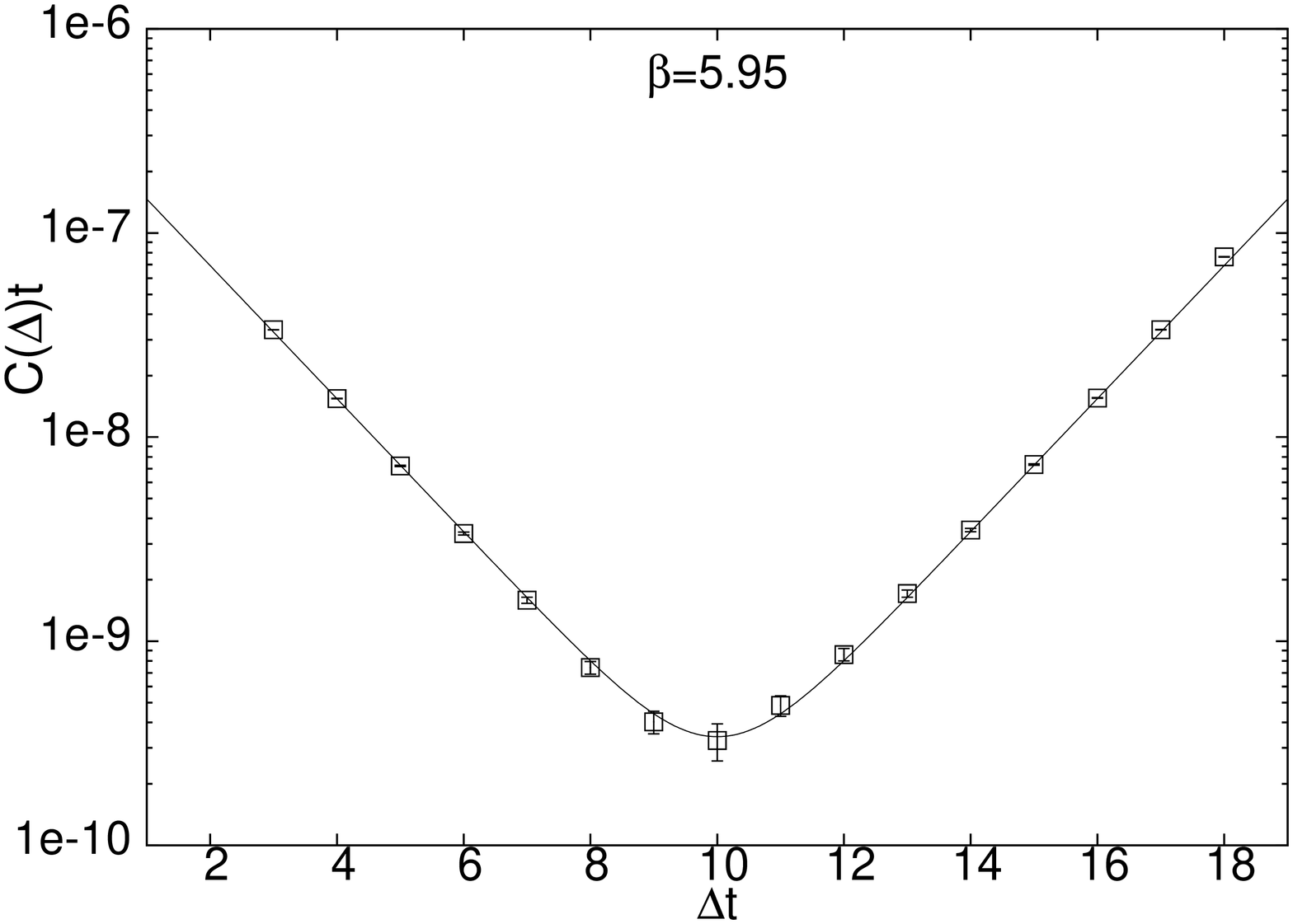}
\includegraphics[width=5.5cm,angle=-90]{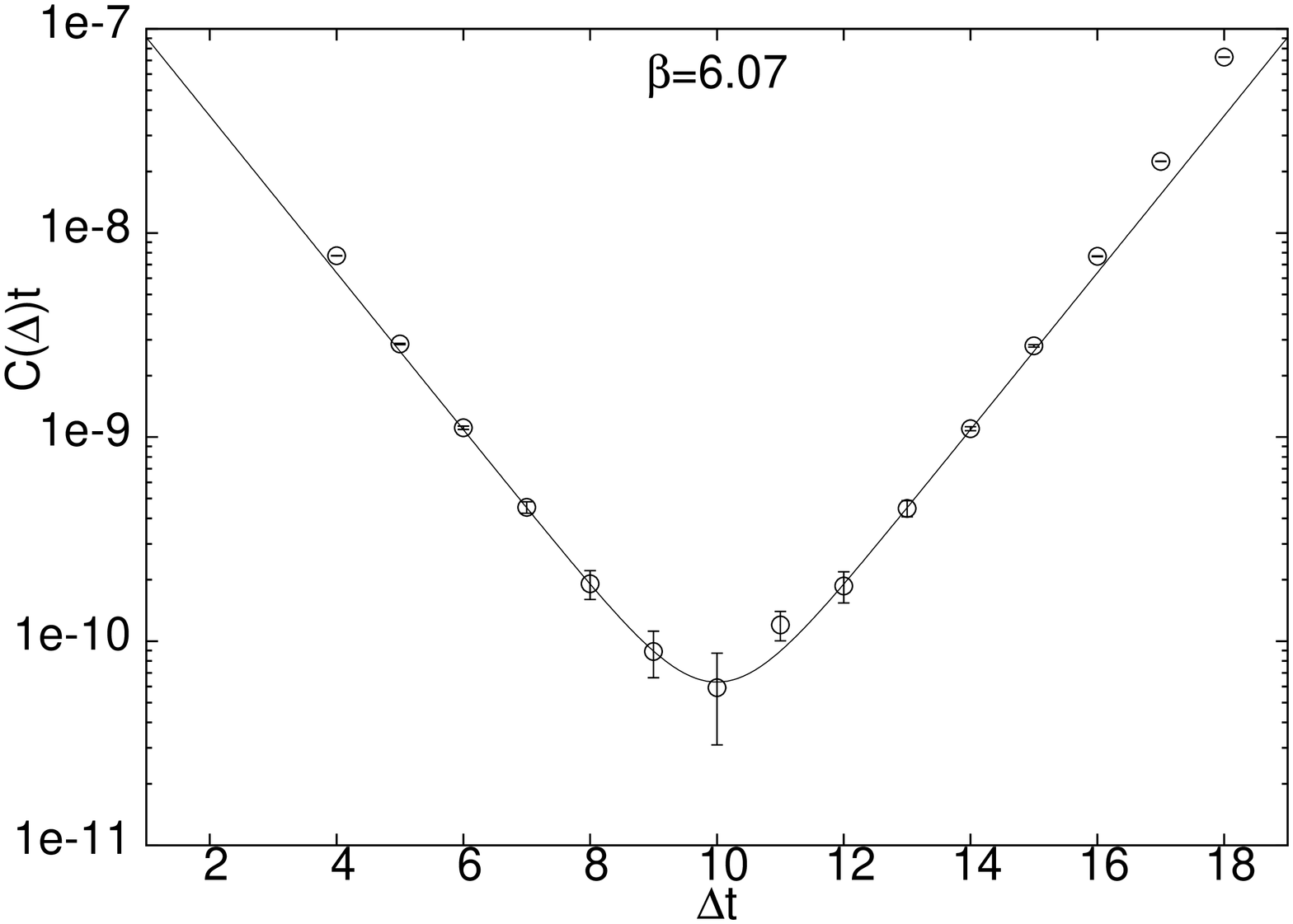}
\caption{\label{fig:mass} The correlators along with their fits at the different $\beta$
values. The left column (with boxes) are for the scalar channel while the right column
(with circles) are for the tensor channel.}
\end{center}
\end{figure*}

\begin{table*}
\begin{center}
\begin{tabular}{|c|c|l|c|c|l|c|c|l|c|c|l|}
\hline
\multicolumn{6}{|c|}{Scalar Channel} & \multicolumn{6}{c|}{Tensor Channel}\\
\hline
&\multicolumn{3}{c|}{global fit}&\multicolumn{2}{c|}{effective mass}& &\multicolumn{3}{c|}{global fit}&\multicolumn{2}{c|}{effective mass} \\\hline
\# & range & ~~~$ma$ & $\frac{{\chi}^2}{d.o.f}$ & t-slice & ~~~$ma$& \# & range & ~~~$ma$ & $\frac{{\chi}^2}{d.o.f}$ & t-slice & ~~~$ma$ \\ \hline
  & 2-9  & ~0.981(3)      & 1.8  & 2/3  & ~0.991(2) &    & 2-7  & ~1.758(9)      & 32.2 & 2/3  & ~1.763(2) \\
  & 3-9  & ~0.961(2)      & 0.05 & 3/4  & ~0.977(6) & D  & 3-7  & ~1.656(12)     & 1.98 & 3/4  & ~1.661(14)\\
A & 4-9  & ~0.962(5)      & 0.06 & 4/5  & ~0.966(22)&    & 4-7  & ~1.585(54)$^*$ & 1.64 & 4/5  & ~1.605(49)\\
  & 5-9  & ~0.952(11)$^*$ & 0.066& 5/6  & ~0.957(41)&    &      &                &      & 5/6  & ~1.39(19) \\
  &      &                &      & 6/7  & ~0.89(12) &    &      &                &      &      &           \\ \hline
  & 2-9  & ~0.936(4)      & 5.7  & 2/3  & ~0.944(1) &    & 4-10 & ~1.166(13)     & 3    & 2/3  & ~1.311(1) \\
  & 3-9  & ~0.915(2)      & 0.3  & 3/4  & ~0.919(4) &  E & 5-10 & ~1.115(39)     & 2.4  & 3/4  & ~1.223(3) \\
B & 4-9  & ~0.904(2)      & 0.05 & 4/5  & ~0.899(8) &    & 6-10 & ~0.938(17)$^*$ & 0.12 & 4/5  & ~1.177(8) \\
  & 5-9  & ~0.911(3)      & 0.025& 5/6  & ~0.909(21)&    &      &                &      & 5/6  & ~1.152(20)\\
  & 6-9  & ~0.906(8)$^*$  & 0.03 & 6/7  & ~0.899(53)&    &      &                &      & 6/7  & ~0.951(52)\\ \hline
  & 3-10 & ~0.765(3)      & 1.3  & 2/3  & ~0.822(1) &    & 4-10 & ~0.988(10)     & 3.3  & 2/3  & ~1.177(1) \\
C & 4-10 & ~0.7537(9)     & 0.04 & 3/4  & ~0.773(2) &  F & 5-10 & ~0.929(10)     & 0.44 & 3/4  & ~1.070(2) \\
  & 5-10 & ~0.7510(15)$^*$& 0.02 & 4/5  & ~0.755(4) &    & 6-10 & ~0.885(16)$^*$ & 0.16 & 4/5  & ~1.004(7) \\
  & 6-10 & ~0.7499(38)    & 0.03 & 5/6  & ~0.751(9) &    &      &                &      & 5/6  & ~0.939(10)\\
  &      &                &      & 6/7  & ~0.734(20)&    &      &                &      & 6/7  & ~0.899(46)\\
  &      &                &      & 7/8  & ~0.723(39)&    &      &                &      & 7/8  & ~0.869(89)\\ \hline
\end{tabular}
\caption{\label{tab:mass} Glueball masses in lattice units ($a$ denotes the lattice spacing)
for all lattices along with the fit parameters. A $^*$ on the mass denotes our best estimate for a
particular coupling and channel.}
\end{center}
\end{table*}

In this section we describe our fitting procedures and the masses we obtain. All the correlators 
 were fitted to the form 
\begin{equation}\label{fitform} 
C(\Delta t)=A\left ( e^{-m\Delta t}+e^{-m(T-\Delta t)}\right )
\end{equation} 
where $m$ is the glueball mass and $T$ is the full temporal extent of the lattice. Since 
the correlator is symmetric about $T/2$, as usual, we fold the data about $T/2$ and use 
only one half of the temporal range for the fits. 

For fitting we use the ``non-linear model fit" of mathematica and  
the fit range was decided on the following two criteria: (i) the range should extend to as 
large a value of $\Delta t$ as possible and (ii) the fit to the form in eq.~\ref{fitform} 
should have a p-value $<$ 0.01 for both $m$ and $A$. We found that the p-value for $A$ gave the 
most stringent criterion for accepting the fit.
The fit range for all the different channels and couplings along with the $\chi^2/d.o.f$ are 
indicated in table~\ref{tab:mass}. 

In addition to masses from global fits, we also compute the effective masses from the correlators as 
\begin{equation}\label{eq:effmass}
am_{\rm eff}=-\ln\frac{\langle C(\Delta t+1)\rangle}{\langle C(\Delta t)\rangle}
\end{equation}
where $a$ is the lattice spacing.
To estimate the error on the effective masses we take $\langle C(\Delta t)\rangle$
to be a jackknife bin and we compute masses for 20 such bins. The error on the effective 
mass is the jackknife error computed from the spread of the masses from the different bins. 
The effective masses are also reported in table~\ref{tab:mass}

In figure \ref{fig:mass} we plot the correlators along with the respective fits for each channel 
and coupling. Even though the fits were done on the 
folded data, in the figures we plot the fitted correlator on the full range. It can be clearly 
seen, especially in the tensor channel that the correlators have contamination form the excited 
states for the smaller values of $\Delta t$. The same thing is seen for the effective masses. 
The masses fall at first and then stabilize to a plateau albeit with increasing error bars for 
larger values of $\Delta t$. 

We cross-check our data by comparing them with results in \cite{LTW,meyer1,meyer2,meyer3,dmgiu}. 
In \cite{LTW} scalar and tensor glueball masses were computed on a symmetric lattice with the 
Wilson action in the $\beta$ range 5.6925 to 6.3380 and we compared mostly with the data presented 
there using exponential interpolation wherever necessary. In the scalar channel at $\beta=5.6925 
~\&~5.6993$ the masses obtained were 0.941(25) and 0.969(18) respectively.    
In \cite{meyer1} the same mass at $\beta=5.7$ was computed to be 0.929(49) and in \cite{dmgiu}, from the 
ratio of partition functions, as 0.935(42). These compare quite well 
with our global fit value of 0.952(11) at $\beta =5.7$. The effective masses we obtain are also 
consitent with this value.
At $\beta =5.7995~\&~5.8$ \cite{LTW} reports values for scalar masses as 0.909(15) and 0.945(21).
We obtain 0.906(8). Finally at $\beta =5.95$ we obtain, using interpolation, a value of 0.743(12) 
from the results in \cite{LTW}. A fit to the correlator gives us a value of 0.7510(15) and our 
effective mass values are also consistent with this estimate.

In the tensor channel at $\beta =5.8$, we look at two lattices ($8^3\times 18$ and $12^3\times 18$)
with different spatial volumes. Unfortunately the data was noisy and we did not get a 
signal for correlators beyond $\Delta t$ of 7. At this $\beta$, we report the results from the 
operator ${\mathcal E}_2$ as the corresponding correlators were less noisy.
For the $8^3\times 18$ lattice we obtain $ma=1.525(35)$ and for the $12^3\times 18$ lattice we get 
$ma=1.585(54)$. This is in the same ball park as the values reported in \cite{LTW} viz. 
$ma=1.52(5)$ at $\beta =5.7995$ and $ma=1.57(6)$ at $\beta =5.8$ both at spatial volumes of $10^3$. 
At $\beta =5.95$, interpolating the data in \cite{LTW} between $\beta =5.8945$ and $\beta =6.0625$ 
we get $ma=1.148(19)$ for the tensor mass. Our best estimate gives $ma=0.938(17)$ for the fit range 
between $\Delta t=6$ and $\Delta t=10$. However if we include the point $\Delta t=5$ in our fit, the 
mass changes to $ma=1.115(39)$. The same trend is there in the effective masses as well.
Between $\Delta t=5 ~\&~ 6$, $am_{\rm eff}$ jumps from around 1.15 to 0.95. 
At $\beta =6.07$, we obtain $ma=0.885(16)$ from our fit and the effective masses are consistent with 
that. The value reported in \cite{LTW} is 0.922(13) at $\beta =6.0625$ and interpolation gives 
$ma=0.913(13)$ at $\beta=6.07$, consistent with our value. At $\beta =5.95~\&~6.07$ our results are from the operator 
${\mathcal E}_1$.

\section{Results - algorithmic gains}\label{adv}
\begin{table}
\begin{center}
\begin{tabular}{|c|c|c|c|c|c|}
\hline
Lattice & Size & $\beta$ & $th$ & iupd & loop size \\ \hline
A$_1$ & ${6^3}\times16$  & 5.7  &2 &20 &$2\times 2$ \\ \hline
B$_1$ & ${6^3}\times18$  & 5.8  &3 &25 &$3\times 3$ \\ \hline
C$_1$ & ${8^3}\times24$  & 5.95 &4 &50 &$5\times 5$ \\ \hline
\hline
D$_1$ & ${6^3}\times18$  & 5.8  &3 &50 &$3\times 3$ \\ \hline
E$_1$ & ${8^3}\times30$  & 5.95 &5 &100&$5\times 5$ \\ \hline
F$_1$ & ${10^3}\times30$ & 6.07 &6 &130&$6\times 6$ \\ \hline
\end{tabular}
\caption{\label{complat} Simulation parameters for additional lattices on which comparisons with the naive method were
carried out. Lattices A$_1$,B$_1$ and C$_1$ were used
for the scalar channel while D$_1$,E$_1$ and F$_1$ were for the tensor channel. ($th$ denotes the sub-lattice thickness)}
\end{center}
\end{table}

To investigate the advantage of the current algorithm over the naive method, we did a few 
 runs for the same computer time using both methods. Since it is not yet clear 
how the algorithm behaves as either $\Delta t$ or $\beta$ changes we report our 
experience for different values of $\Delta t$ and $\beta$.  

For the lattice D$_1$, we carried out runs for 200 hours. The multilevel 
algorithm had an error of 3\% at $\Delta t=3$ which is just below $r_0$ (see table~\ref{lattices}), 
while the naive algorithm had an error of 81\%. It would be interesting to compare the 
performance at a value of $\Delta t$ between $r_0$ and $2r_0$. So we choose points around $1.5r_0$
(in this case $\Delta t=6$).
Even after 200 hrs of runtime we do not have a signal at that distance for the naive method. So to estimate 
the \% error we multiply the naive correlator at the largest value of $\Delta t$ where we 
have a signal (i.e. $\Delta t=3$) by 
${\rm corr}_{\rm multilevel}(\Delta t=6)/{\rm corr}_{\rm multilevel}(\Delta t=3).$ 
Doing this we get the \% error to be 850\% for the naive method while it is 29\% 
for the multilevel scheme. Thus for the tensor channel at $\beta=5.8$ we estimate 
that the error reduction algorithm produces an error which is between 27 times smaller 
than the naive method at both $\Delta t = 3 {\rm ~and~} 6 $. Since the error 
$\propto$ time$^2$ we estimate the new method is more efficient by at least a factor 
of 729 or so. 

At $\beta=5.95$ (lattice E$_1$), the runs were for about 100 hours. There at $\Delta t=3$ the multilevel 
algorithm produced an error of about 4\% while the naive algorithm had an error of 75\%.
Doing a similar estimate as $\beta=5.8$ we estimate that at $1.5r_0$ ($\Delta t=8$) the 
errors are 150\% for the multilevel algorithm while it is about 3000\% for the naive 
method. At this $\beta$ value therefore, the gains in terms of \% error is about 
a factor 20. At $\beta=6.07$, we did not get a signal for the naive algorithm for any $\Delta t$
other than $\Delta t=1$ even after about 300 hours of runs. Thus we see that the gain has very 
little dependence on $\Delta t$ but does depend on $\beta$.

\begin{table}
\begin{center}
\begin{tabular}{|c|c|c|c|c|c|c|c|}
\hline
\multicolumn{4}{|c}{Scalar channel} & \multicolumn{4}{|c|}{Tensor channel}\\
\hline
\# & time & \vspace*{-1mm}${\rm err}_{\rm n}$ & $gain$ & \# & time & ${\rm err}_{\rm n}$ & $gain$\\
& (min) & $\overline{{\rm err}_{\rm ml}}$ &  & & (min) & $\overline{{\rm err}_{\rm ml}}$ & \\
\hline
A & 3850 & 5.7 & 32 & D$_1$ & 12000 & 27 & 729 \\
B$_1$ & 1000 & 5.5 & 30 & E$_1$ & 5775 & 20 & 400 \\
C$_1$ & 1100 & 18 & 324 & F$_1$ & 15000 & $-$ & $-$  \\
\hline
\end{tabular}
\caption{\label{tab:comp} Comparison of error bars between the naive and error reduction
methods. Please see the text in section~\ref{adv} for a discussion on these values. ${\rm err}_{\rm n}$
stands for error in the naive method while ${\rm err}_{\rm ml}$ denotes error in the multilevel scheme.
$gain$ is in terms of time and is given by $({\rm err}_{\rm n}/{\rm err}_{\rm ml})^2$.}
\end{center}
\end{table}

For the scalar channel using lattice A, runs were carried out for about 3850 mins. Comparing the 
errors around $1.4r_0$, we got a gain of about 5.7 in terms of errors or 32 in terms of time.
At $\beta=5.8$ (lattice B$_1$) the runs were carried out for about 1000 mins. In this case 
we have a signal at $1.5r_0$ for both methods and we get an error of 13\% for the multilevel scheme 
while it is about 70\% for the naive method. Thus the gain in terms of \% error is about 5.5 or in 
terms of time about 30. At $\beta=5.95$ in the scalar channel, again we do not have a signal at 
$1.5r_0$ using the naive method and are forced to use the same method as in the tensor channel to 
estimate the errors. At $\Delta t=3$ we obtain the errors to be 2\% and 37\% for the multilevel and 
the naive methods respectively while at $1.5r_0$ they are 29\% and 500\% (estimated) respectively. 
Thus the ratio of errors is about 18 or gain in terms of time 324. 
A compilation of our results is presented in table~\ref{tab:comp}.

In addition to the above, at $\beta=5.7$ we have one more comparison using the lattice A$_1$.
There we obtain a gain of 2.5 in terms of errors or 6 in terms of time. Thus the gain seems to increase 
with increase in volume. We expect this will help us go to larger lattices.

Error reduction techniques only reduce statistical errors. There are systematic errors as well 
and the most important among that are finite volume effects. In our 
lattices with small physical volumes (B$_1$ to F$_1$), we encounter them. For example 
at $\beta=5.8$ (lattices B$_1$ and D$_1$) the mass in the tensor channel is smaller than the mass in 
the scalar channel which is the expected behaviour at small volumes~\cite{anafvol,su2fvol,su3fvol}. 
For a recent study of finite volume effects we refer the reader to \cite{fvol}.
To mitigate these we choose our lattices (A to F) such that $mL>9$ in all cases~\cite{MM}.

\section{Conclusions}
 Extraction of glueball masses from correlators is a difficult problem in lattice QCD due to a 
very low signal to noise ratio at large Euclidean times. In this article we present a new method, 
based on the multilevel scheme, to enhance the signal to noise ratio in glueball correators. 
We observe that this error reduction technique works quite well at least in pure gauge theories. 
For a given computational cost, the improvement over the naive method in the signal to noise ratio 
is several times to more than an order of magnitude. We are able to follow the correlator to temporal 
separations of about 1 fermi and can perform global fits to the correlators between 0.5 and 1 fermi. 
Our effective masses also show a plateau in the same range obtained from the global fits. 

We improve upon the existing error bars on the masses in the scalar channel and 
in the tensor channel our error bars are comparable to the existing ones in the range of $\beta$ that 
we have looked at.  
It is of course of interest to reach the continuum limit and we are continuing our runs at finer and 
larger lattices  and will report our results in subsequent publications.

\section*{Acknowledgments}
The runs were carried out partly on the cluster bought under the DST project SR/S2/HEP-35/2008 
and partly on the CRAY XE6-XK6 at IACS. The authors would like to thank DST, IACS and ILGTI for these 
facilities. The authors would also like to thank Peter Weisz for a careful reading of the manuscript 
and his comments on finite volume effects.

\end{document}